\documentstyle[12pt]{article}
         \makeatletter
         \def\thefigure{\@arabic\c@figure}\def\fps@figure{tbp}
         \def\ftype@figure{1}\def\ext@figure{lof}
         \def\fnum@figure{\protect\footnotesize Fig.\ \thefigure}
         \def\thetable{\@arabic\c@table}
         \def\fps@table{tbp}\def\ftype@table{2}\def\ext@table{lot}
         \def\fnum@table{\protect\footnotesize Table \thetable}
         \def\@listI{\leftmargin\leftmargini\parsep=0pt\itemsep=0pt}
         \def\thebibliography#1{\section{References}\vspace*{-10pt}\list
          {[\arabic{enumi}]}{\settowidth\labelwidth{[#1]}\leftmargin\labelwidth
          \advance\leftmargin\labelsep
          \usecounter{enumi}}
          \def\newblock{\hskip .11em plus .33em minus .07em}
          \sloppy\clubpenalty4000\widowpenalty4000
          \sfcode`\.=1000\relax}
         
         \def\@nomath#1{\ifmmode \fi}
         \def\mmycite{\@ifnextchar [{\@tempswatrue\@mmycitex}
             {\@tempswafalse\@mmycitex[]}}
         \def\@mmycitex[#1]#2{\if@filesw\immediate%
         \write\@auxout{\string\citation{#2}}\fi
           \def\@citea{}\@mmycite{\@for\@citeb:=#2\do
             {\@citea\def\@citea{,}\@ifundefined
                {b@\@citeb}{{\bf ?}\@warning
                {Citation `\@citeb' on page \thepage \space undefined}}%
         \hbox{\csname b@\@citeb\endcsname}}}{#1}}
         \def\@mmycite#1#2{{{\scriptsize#1}\if@tempswa , #2\fi}}
         \def\mycite#1{$^{\protect\mmycite{#1}}$}
         \def\@cite#1#2{{#1\if@tempswa , #2\fi}}
         \def\thesection {\arabic{section}}
         \def\section#1{\addtocounter{section}{1}\setcounter{subsection}{0}
              \bigskip\medskip{\noindent\bf\thesection.\ #1}
              \medskip}
         \def\thesubsection {\arabic{section}.\arabic{subsection}}
         \def\subsection#1{\addtocounter{subsection}{1}
              \medskip{\noindent\thesubsection.\ #1}
              \medskip}
         \makeatother
\def\bit{\begin{itemize}}
\def\eit{\end{itemize}}

\def\part{{\partial_t}}

\def\au#1 {\begin{center} #1 \end{center}}
\def\case#1/#2{{\textstyle\frac{#1}{#2}}}
\def\cntr#1 {\begin{center} #1 \end{center}}
\def\eq#1 #2 {\begin{equation} \label{#1} #2 \end{equation}}
\def\eqa#1 #2 #3 {\begin{eqnarray} \label{#1} #2 \label{#3} \end{eqnarray}}

\def\fig#1 #2 #3 #4 {\begin{figure} \vspace{#3pt} \caption[#1]{#4} \label{#1}
\end{figure}}

\def\tbl#1 #2 #3 #4 {\begin{table} \caption[#1]{#3} \label{#1} \vspace{-6pt}
\begin{center} {\begin{tabular}{#2}  \hline\hline #4 \vspace{1pt} \\
\hline\hline \end{tabular}} \end{center} \end{table}}

\def\tblb#1 #2 #3 #4 {\begin{table}[b] \caption[#1]{#3} \label{#1}
\vspace{-6pt} \begin{center} {\begin{tabular}{#2}  \hline\hline #4 \vspace{1pt}
\\ \hline\hline \end{tabular}} \end{center} \end{table}}

\def\tblh#1 #2 #3 #4 {\begin{table}[h] \caption[#1]{#3} \label{#1}
\vspace{-6pt} \begin{center} {\begin{tabular}{#2}  \hline\hline #4 \vspace{1pt}
\\ \hline\hline \end{tabular}} \end{center} \end{table}}

\def\ti#1 {\begin{center} \baselineskip=17pt {\large #1} \end{center}}

\def\tibf#1 {\begin{center} \baselineskip=17.5pt {\large \bf #1} \end{center}}

\def\ess{\hskip.444444em plus .499997em minus .037036em}
\def\mss{\hskip.333333em plus .208331em minus .088889em}

\def\sen{\hbox{\scriptsize--}}
\def\eV{e\kern-.10emV }
\def\eVcm{e\kern-.10emV\kern-.15em,\mss}
\def\eVcl{e\kern-.10emV\kern-.10em:\ess}
\def\eVsc{e\kern-.10emV\kern-.10em;\mss}
\def\eVp{e\kern-.10emV\kern-.15em.\ess}
\def\eVpr{e\kern-.10emV) }
\def\eVc{e\kern-.10emV\kern-.10em/\kern-.10em$c$ }
\def\eVccm{e\kern-.10emV\kern-.10em/\kern-.10em$c$, }
\def\eVcp{e\kern-.10emV\kern-.10em/\kern-.10em$c$. }
\def\eVf{e\kern-.10emV\kern-.10em/fm }
\def\eVfcm{e\kern-.10emV\kern-.10em/fm, }
\def\eVfp{e\kern-.10emV\kern-.10em/fm. }

\begin{document}

\vspace*{0.3in}
\begin{center}
  {\bf QUANTUM SIZE EFFECTS IN CLASSICAL HADRODYNAMICS}\\
  \bigskip
  \bigskip
  J. Rayford NIX\\
  {\em Theoretical Division, Los Alamos National Laboratory\\
       Los Alamos, New Mexico 87545, USA}\\
  \bigskip
\end{center}
\smallskip
{\footnotesize
\centerline{ABSTRACT}
\begin{quotation}
\vspace{-0.10in}
We discuss future directions in the development of classical hadrodynamics for
extended nucleons, corresponding to nucleons of finite size interacting with
massive meson fields.  This new theory provides a natural covariant microscopic
approach to relativistic nucleus-nucleus collisions that includes automatically
spacetime nonlocality and retardation, nonequilibrium phenomena, interactions
among all nucleons, and particle production.  The present version of our theory
includes only the neutral scalar ($\sigma$) and neutral vector ($\omega$) meson
fields.  In the future, additional isovector pseudoscalar
($\pi^+$,~$\pi^-$,~$\pi^0$), isovector vector ($\rho^+$,~$\rho^-$,~$\rho^0$),
and neutral pseudoscalar ($\eta$) meson fields should be incorporated.  Quantum
size effects should be included in the equations of motion by use of the
spreading function of Moniz and Sharp, which generates an effective nucleon
mass density smeared out over a Compton wavelength.  However, unlike the
situation in electrodynamics, the Compton wavelength of the nucleon is small
compared to its radius, so that effects due to the intrinsic size of the
nucleon dominate.
\end{quotation}}

\section{Introduction}

During three previous winter workshops we have discussed a new microscopic
many-body approach to relativistic nucleus-nucleus collisions based on
classical hadrodynamics for extended nucleons, corresponding to nucleons of
finite size interacting with massive meson
fields.\mycite{SHN90}$^{\sen}$\mycite{BN92}\ess  The underlying foundations of
this new theory, as well as applications to soft nucleon-nucleon collisions,
have been published recently.\mycite{BN93a,BN93b}\ess  In this contribution, we
would like to discuss future directions in the systematic development of this
theory.

\section{Extended Nucleon}

We all know that the nucleon is a composite particle made up of three valence
quarks plus additional sea quarks and gluons.  When nucleons collide at very
high energies, a few rare events correspond to the head-on or hard collisions
between the individual quarks and/or gluons.  For describing these events, the
underlying quark-gluon structure of the nucleon is of crucial importance.  Yet
such hard collisions are extremely rare, typically one in a
billion.\mycite{Pe87}\ess  The vast majority of events correspond to soft
collisions not involving individual quarks or gluons.  For the description of
such events, an appealing idea is to regard the nucleon as a single extended
object interacting with other nucleons through the conventional exchange of
mesons (whose underlying quark-antiquark composition is ignored).

Experiments involving elastic electron scattering off protons have determined
that the proton charge density is approximately exponential in
shape,\mycite{SBSW80}\mss with a root-mean-square radius of 0.862 $\pm$ 0.012
fm.  Although many questions remain concerning the relationship between the
proton charge density and the nucleon mass density,\mycite{Bh88}\mss it should
be a fairly accurate approximation to regard them as equal.  We therefore take
the nucleon mass density to be

\eq  rhonuc {\rho(r) = \frac{\mu^3}{8\pi} \exp(-\mu r) ~,}
with $\mu = \sqrt{12}/R_{\rm rms}$ and $R_{\rm rms}$ = 0.862 fm.  We show in
fig.~\ref{nucleonf} a gray-scale plot of the mass density through the center of
a nucleon calculated according to this exponential, with three valence quarks
also indicated schematically.

\fig nucleonf snowbird/nucleon 267 {\footnotesize Slice through the center of a
nucleon.  Although composed of the three indicated valence quarks plus sea
quarks and gluons, for many purposes the nucleon can be regarded as a single
extended object, with an exponentially decreasing mass density that is
spherically symmetric in its instantaneous rest frame.}

\section{Present Version}

In the present version of our theory, we consider $N$ extended, unexcited
nucleons interacting with massive, neutral scalar ($\sigma$) and neutral vector
($\omega$) meson fields.  At bombarding energies of many G\eV per nucleon, the
de Broglie wavelength of projectile nucleons is extremely small compared to all
other length scales in the problem.  Furthermore, the Compton wavelength of the
nucleon is small compared to its radius, so that effects due to the intrinsic
size of the nucleon dominate those due to quantum uncertainty.  The classical
approximation for nucleon trajectories should therefore be valid, provided that
the effect of the finite nucleon size on the equations of motion is taken into
account.  The resulting classical relativistic many-body equations of motion
can be written as\mycite{BN93a,BN93b}

\begin{equation}
M^*_i a_i^\mu = f_{{\rm s},i}^\mu + f_{{\rm v},i}^\mu + f_{{\rm s,ext},i}^\mu +
f_{{\rm v,ext},i}^\mu ~~~,
\end{equation}
where $M^*_i$ is the effective mass, $f_{{\rm s},i}^\mu$ is the scalar
self-force, $f_{{\rm v},i}^\mu$ is the vector self-force, $f_{{\rm
s,ext},i}^\mu$ is the scalar external force, and $f_{{\rm v,ext},i}^\mu$ is the
vector external force.

These classical relativistic many-body equations of motion can be solved
numerically without further approximation.  In particular, there is
\bit
\item No mean-field approximation
\item No perturbative expansion in coupling strength
\item No superposition of two-body collisions
\eit
We have thus far solved these equations for soft nucleon-nucleon collisions
at $p_{\rm lab}$ = 14.6, 30, 60, 100, and 200 G\eVc to yield such physically
observable quantities as scattering angle, transverse energy, radiated energy,
and rapidity.\mycite{BN93b}  We found that the theory provides a physically
reasonable description of gross features associated with the soft reactions
that dominate nucleon-nucleon collisions.  In addition, the present version of
the theory permits a qualitative discussion of several important physical
points:
\bit
\item Effect of finite nucleon size on equations of motion
\item Inherent spacetime nonlocality
\item Particle production through massive bremsstrahlung
\eit
\vspace{-10pt}

\section{Additional Meson Fields}

Nevertheless, from nucleon-nucleon scattering experiments we know that several
additional meson fields are important and must be included for a quantitative
description:\mycite{Ma89}

\bit
\item Isovector pseudoscalar ($\pi^+$, $\pi^-$, $\pi^0$)
\item Isovector vector ($\rho^+$, $\rho^-$, $\rho^0$)
\item Neutral pseudoscalar ($\eta$)
\eit
The next step in the systematic development of the theory should be the
inclusion of these additional meson fields.

\newpage

\fig rhoeffe snowbird/rhoeffe2 238 {\footnotesize Effective charge density for
a point electron.  The intrinsic charge density is the positive delta function
indicated by the vertical dashed line at the origin.}

\section{Quantum Spreading Function}

Although effects due to the intrinsic size of
the nucleon dominate those due to quantum uncertainty, it is nevertheless
important to provide an estimate of the effects of quantum uncertainty and to
include them in the equations of motion if they are important.  This should be
possible by use of techniques analogous to those used by Moniz and
Sharp\mycite{MS77} for nonrelativistic quantum electrodynamics.

\subsection{Nonrelativistic Quantum Electrodynamics}

There appear in the classical nonrelativistic equations of motion for an
extended electron terms of the form $\int_{\infty} \int_{\infty} \rho(r) \,
{\cal O} \rho(r') \, d^3 \! r \, d^3 \! r'$, where the operator ${\cal O}$ is a
function of ${\bf r}$ and ${\bf r'}$.  Moniz and Sharp\mycite{MS77} have shown
in nonrelativistic quantum electrodynamics that the effect of quantum mechanics
on the equations of motion is to replace such terms by terms of the form
$\int_{\infty} \int_{\infty} \rho(r) \, {\cal O_{\rm \! eff}} \, \rho_{\rm
eff}(r') \, d^3 \! r \, d^3 \! r'$ and derivatives with respect to $\lambda$ of
these terms.  The effective operator ${\cal O_{\rm \! eff}}$ is a function of
${\bf r}$, ${\bf r'}$, and $\lambda^2 \, {\bf {\nabla\kern-.15em_{\bf
r'}}^2}$.  The effective density $\rho_{\rm eff}$ is given by

\eq rhoeff {\rho_{\rm eff}(r) = \int_{\infty} S(|{\bf r}-{\bf r'}|) \, \rho(r')
\, d^3 \! r' ~,}
where

\eq S {S(|{\bf r}-{\bf r'}|) = -\frac{\cos(2|{\bf r}-{\bf
r'}|/\lambda)}{\pi\lambda^2 \, |{\bf r}-{\bf r'}|}}
and $\lambda = 1/(M_0)$ is the Compton wavelength associated with the
particle's bare mass.  [Please note that we have made some obvious corrections
to eq.~(3.27) of ref.~\cite{MS77}.]

For a point electron, where the intrinsic charge density is $\rho_{\rm e}(r) =
\delta_+(r)/(4\pi r^2)$, eq.~(\ref{rhoeff}) reduces to

\eq electron {\rho_{\rm eff}(r) = -\frac{\cos(2r/\lambda_{\rm
e})}{\pi{\lambda_{\rm e}}^2 \, r} ~,}
with a Compton wavelength $\lambda_{\rm e}$ = 386.15933 fm.  As illustrated in
fig.~\ref{rhoeffe}, the effective charge density for a point electron is a
decreasing oscillatory function of radial distance from the origin.

\subsection{Classical Hadrodynamics}

For an extended nucleon, where the intrinsic mass density is given by the
exponential~(\ref{rhonuc}), eq.~(\ref{rhoeff}) leads to

\eq nucleon {\rho_{\rm eff}(r) = \frac{\mu^3}{8\pi\lambda \, r} \int_0^{\infty}
r' \exp(-\mu r') \, \{\sin(2|r-r'|/\lambda) - \sin[2(r+r')/\lambda]\}  \, dr'
{}~.}
For a bare nucleon mass\mycite{BN93a} $M_0$ = 949.47 M\eVcm the Compton
wavelength $\lambda$ = 0.20783~fm, which is much smaller than the
root-mean-square radius.  This has the consequence, as illustrated in
fig.~\ref{rhoeff}, that the effective mass density for an extended nucleon
oscillates around the intrinsic mass density as a function of radial distance
from the origin.

\fig rhoefff snowbird/rhoeff2 238 {\footnotesize Effective mass density for an
extended nucleon.  The intrinsic mass density is exponential with a
root-mean-square radius of 0.862~fm.}

\newpage

\section{Conclusions}

To conclude, we have shown that classical hadrodynamics
\bit
\item Provides a manifestly Lorentz-covariant microscopic many-body
approach to relativistic heavy-ion collisions
\item Satisfies {\it a priori\/} the basic conditions that are present
\item Requires minimal physical input
\item Leads to equations of motion that can be solved numerically without
further approximation
\item Contains an inherent spacetime nonlocality that may be responsible for
significant collective effects
\item Provides in its present form a qualitative description of transverse
momentum, radiated energy, and other gross features in nucleon-nucleon
collisions
\item Should be extended to include additional meson fields for the $\pi$,
$\rho$, and $\eta$, as well as quantum size effects
\eit
\vspace{-10pt}

\section{Acknowledgments}

We are grateful to R.~J. Hughes and A.~J. Sierk for their participation during
the early stages of this investigation, to B.~W. Bush for his pioneering
contributions during later stages, and to D.~H. Sharp for a stimulating
discussion concerning his equations and their interpretation.  This work was
supported by the U.~S. Department of Energy.

\end{document}